\documentclass[12pt,preprint]{aastex}

\shorttitle{DISCOVERY OF AN ULTRACOOL WHITE DWARF COMPANION}
\shortauthors{J. Farihi}

\begin{document}

\title{DISCOVERY OF AN ULTRACOOL WHITE DWARF COMPANION}

\author{J. Farihi}
\affil{Department of Physics \& Astronomy, 8371 Math Sciences Building, 
	University of California, Los Angeles, CA 90095-1562}
\email{jfarihi@astro.ucla.edu}

\begin{abstract}

The discovery of a low luminosity common proper motion companion
to the white dwarf GD392 at a wide separation of $46''$ is reported.
$BVRI$ photometry suggests a low temperature ($T_{\rm eff}\sim4000$
K) while $JHK$ data strongly indicate suppressed flux at all near
infrared  wavelengths.  Thus, GD392B is one of the few white dwarfs 
to show significant collision induced absorption due to the presence
of photospheric ${\rm {H_2}}$ and the first ultracool white dwarf
detected as a companion to another star.  Models fail to explain
GD392B as a normal mass white dwarf.  If correct, the cool companion
may be explained as a low mass white dwarf or unresolved double
degenerate.  The similarities of GD392B to known ultracool
degenerates are discussed, including some possible implications for
the faint end of the white dwarf luminosity function.

\end{abstract}


\section{INTRODUCTION}

White dwarfs are evolved, degenerate stars which have
exhausted their nuclear fuel and are supported by electron
degeneracy pressure.  They are destined to cool to ever lower
temperatures over billions of years as the end product of most
stars in the Galaxy, main sequence stars with $M\la8M_{\odot}$.
Next to low mass stars, they are the most common stellar objects
in the Milky Way, providing a window into the history of stellar
evolution and star formation in the Galaxy.
	
Ultracool white dwarfs ($T_{\rm eff}<4000$ K) have been
of much interest to researchers in recent years for several
reasons.  First, it remains an unanswered question as to
whether cool white dwarfs are a significant component of
Galactic halo dark matter \citep{han03}.  Second, the ages
and compositions of cool white dwarfs have obvious implications
for the age and evolution of the disk, bulge and halo components
of the Galaxy.  Finally, the distribution and origins of
atmospheric and core compositions of the coolest white dwarfs
is still not well understood.

In this paper the discovery of GD392B, a common proper motion
companion to the helium white dwarf GD392 (WD2058+342, $21^{\rm
h} 00^{\rm m} 21.5^{\rm s}, +34\arcdeg 26' 20''$ J2000), is
discussed.  The spectral energy distribution (SED) of GD392B
strongly suggests flux deficiency redward of $1\mu$m, a $T_{\rm
eff}<4000$ K, and possibly a helium component in its atmosphere
as well \citep{ber02}.  Tentative stellar parameters are derived
for GD392B based on the data and different assumptions about the
mass of the primary.  Possible implications based on the model
fit parameters and similarities to known ultracool white
dwarfs are then discussed.

\section{OBSERVATIONS \& DATA REDUCTION}

\subsection{\it Photometry}

Near infrared data on the GD392 system were obtained at Lick
Observatory in June \& August 2003.  $JHK^{'}$ images were
acquired using the Gemini camera \citep{mcl93} on the 3 meter
Shane telescope.  Total observation time was 7.5 minutes at each
filter, consisting of 15 second exposures in a repeated five point
dither pattern.  Multiple infrared photometric standard stars from
\citet{hun98} \& \citet{haw01} were observed twice during the night.
The conditions in August were very good with clear skies and average
seeing of $1.3''$.  In June the weather was good but the average
seeing was $1.8''$.  A few $K$ band images were taken at Mauna Kea
in July 2003 using the Near Infrared Camera (NIRC) \citep{mat94}
on the 10 meter Keck I telescope.  Images consisted of 10 second
exposures for the primary and 50 seconds for the secondary.  Observing
conditions were poor with considerable cloud cover and seeing of
$0.6''$.  Optical data were obtained at Lick Observatory in July \& 
September 2003.  $BVRI$ images were acquired with the CCD camera on
the 1 meter Nickel telescope.  Total exposure time was 5 minutes at 
each filter and photometric standard stars from \citet{lan83} were 
observed immediately prior to the GD392 system.  Conditions in August 
were excellent with clear skies and seeing of $0.9''$.  In July, the 
sky was clear but the average seeing was $1.2''$

Images were reduced using standard programs in the IRAF
environment.  Optical images were filtered clean of bad pixels
and cosmic rays in the area of interest, then flat fielded with
a normalized dome flat.  Near infrared images were sky subtracted,
flat fielded, registered and averaged into a single image
at each wavelength.

Straightforward aperture photometry was used, including
airmass/extinction corrections, for $BVRIJHK$ measurements of
the primary.  Using a circular aperture centered on the target
star and an annulus on the surrounding sky, both the flux and
signal to noise ratio (SNR) was calculated for a range of
apertures from one to four full widths.  The target flux was
measured at or near the aperture size which produced the
largest SNR.  In this way the flux of all targets was measured,
including standard stars.  A large aperture ($d\sim6''$) was used
for all calibrators, and flux measurements for science targets
were corrected to this standard aperture to account for the
smaller percentage of the total flux contained within apertures
smaller than the standard.  

\begin{figure}

\plotone{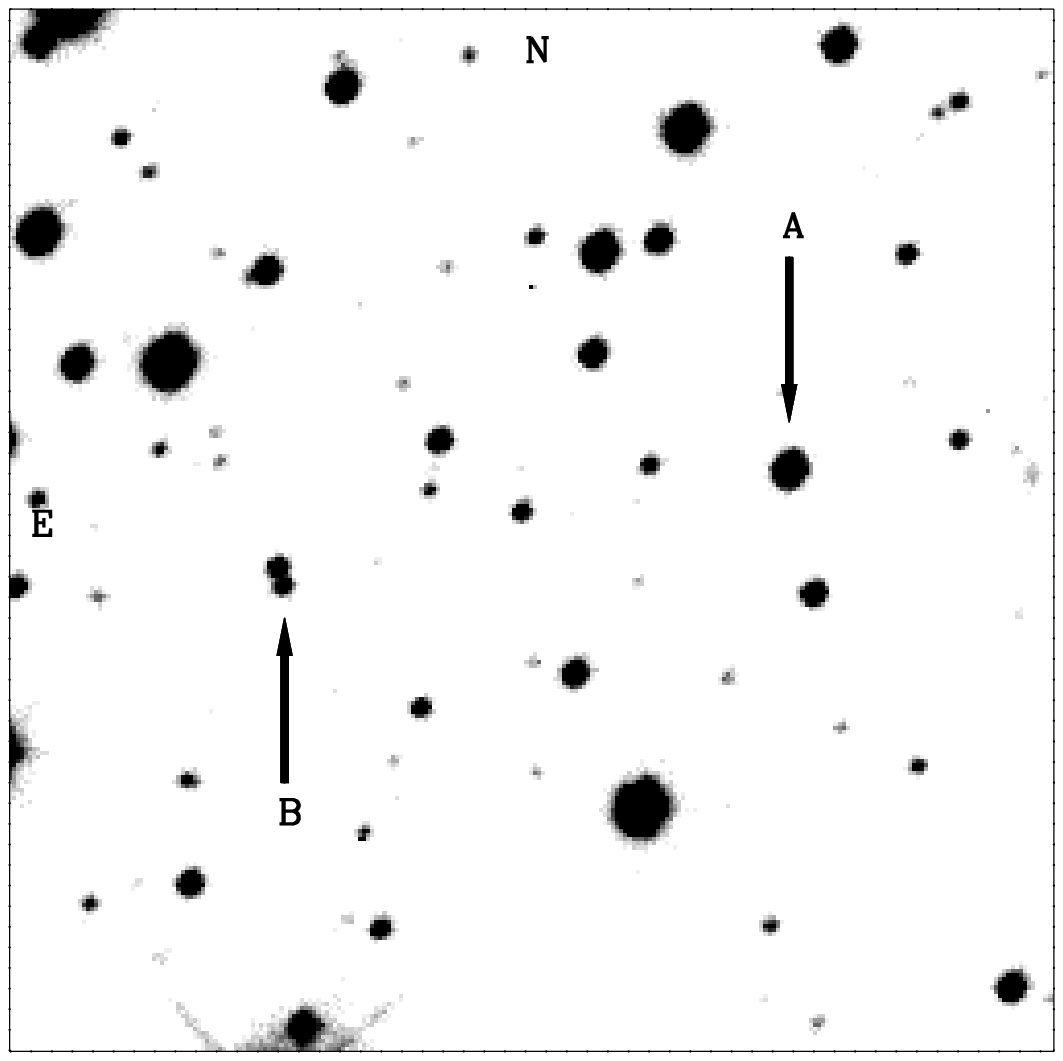}
\caption{$I$ band ($0.80\mu$m) image of the GD392 system
obtained with the CCD camera on the Nickel 1 m telescope on
2003 September 15.  The scale is $0.36''$ per pixel and the
frame is $92''$ on a side.  The object labelled `A' is GD392
and the object labelled `B' is the cool white dwarf companion,
GD392B ($21^{\rm h} 00^{\rm m} 25.1^{\rm s}, +34\arcdeg 26' 09''$
J2000).  About $1.8''$ NE of the secondary is an unrelated
background K dwarf.  Due to its proper motion, GD392B and the
background K star are moving closer together.
\label{fig1}}
\end{figure}

Extracting all photometry for GD392B was complicated by the
presence of a nearby background star (Figure \ref{fig1}).
In a single subtraction $J$ band image, the separation between
GD392B and background star was measured to be $1.8\pm0.1''$ at
a PA of $196\pm3\arcdeg$.  Since even small aperture flux
measurements of GD392B would be contaminated by light from the
background star, the IRAF package DAOPHOT was employed to
fit the PSF of both stars simultaneously and extract their
magnitudes.  Since DAOPHOT is sensitive to the separation and
brightness ratio of two sources with overlapping PSFs, the
optical and infrared data were broken into two sets; primary
data for which the full widths at half maximum in the reduced
images were less than $1.8''$, and secondary data for which they
were greater than or equal to $1.8''$.  This is an especially
important consideration in the infrared where the brightness
ratio of GD392B and background star is far from unity.  All 
photometric data are listed in Table \ref{tbl-1} and the adopted
values used for the analysis are listed in Table \ref{tbl-2}.
The secondary data were not included in the analysis due to the
uncertainty in PSF fitting for sources separated by less than
one full width at half maximum -- the data are consistent with
the adopted values to within $1-2 \sigma$ and are included for
completeness.  Conversions between near infrared filter sets 
were not performed for three reasons: 1) the corrections are 
typically very small \citep{leg92,car01}; 2) the colors of GD392B 
are drastically different from the colors of standard stars used 
to derive the transformations; 3) the near infrared photometry 
errors in Table \ref{tbl-1} are likely to be several times larger
than any corrections.

Due to the poor weather conditions, there was only a single NIRC
frame acquired with GD392B on the chip, three frames containing
GD392A, and several sky frames.  Since the common proper motion
system is separated by more than the $39''$ field of view, the 
binary pair could not be observed simultaneously.  The reduced 
images of the primary and secondary contained three field stars 
in common.  Aperture photometry was performed on these stars twice;
once with GD392A as a calibrator, and again using the three field 
stars as calibrators for GD392B and the background star.  These 
intermediate calibrator stars had a standard deviation of only a 
few percent in their measured relative fluxes between the two 
reduced images.  GD392B and the background star were spatially 
resolved from each other in this observation.

The SNR in the primary data set was greater than 30 for both
stars in the reduced images at all wavelengths with the exception
of the following.  In the Gemini images, the SNR was calculated to
be 8.6 at $H$, 3.7 at $K^{'}$ for GD392B (from the extracted
DAOPHOT data).  In the Nickel data, the SNR at $B$ was 11.4 for
GD392B and 15.1 for the background star.   In the single
subtraction, flat fielded NIRC $K$ frame, the SNR for GD392B
was 5.8.

\subsection{\it Spectroscopy}

A $0.38-1.0\mu$m spectrum was obtained using the Low
Resolution Imaging Spectrograph (LRIS) \citep{oke95} in October
2003 at Mauna Kea on Keck I by C. Steidel, D. Erb \& N. Reddy.  
The setup consisted of the 300/5000 \AA{} grism on the blue side 
and the 400/8500 \AA{} grating on the red side, separated at 6650 
\AA{} by a dichroic beamsplitter. The spectra of both stars were 
spatially resolved from each other in the observation.  The spectral 
images were cleaned of bad pixels and cosmic rays, then flat fielded 
using an internal halogen lamp.  A spectrum of the sky was extracted 
at two positions for each wavelength region, averaged, and then
subtracted from the spectrum of each star.  Standard programs in
IRAF were used to extract the spectra of both GD392B and the
background star.

\begin{figure}

\plotone{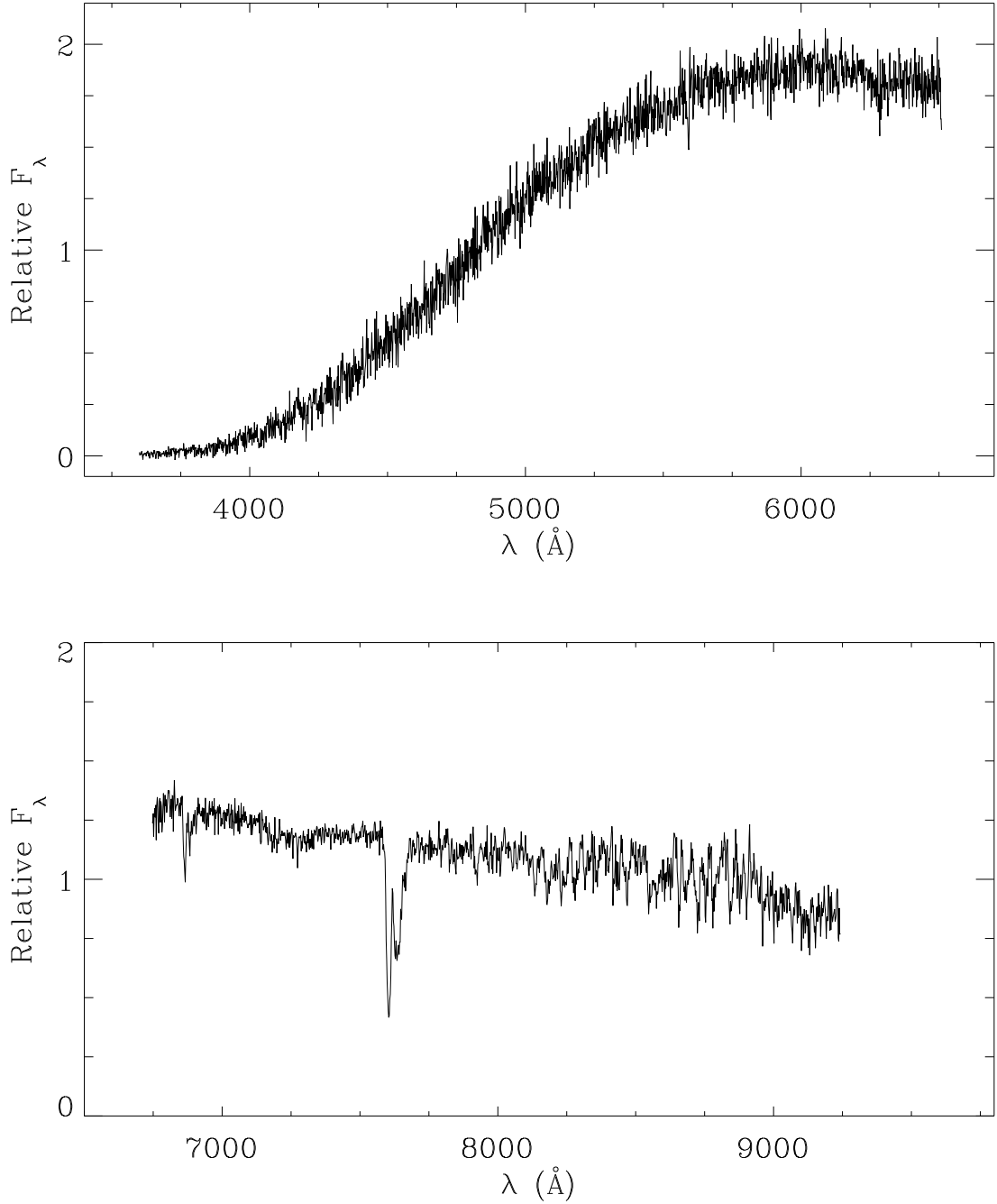}
\caption{Optical spectrum of GD392B taken with LRIS.  The
data are not flux calibrated but the overall shape should be
fairly accurate (Figure \ref{fig3}).  The only real features
are those at 6870 \& 7590 \AA{} and are due to telluric
${\rm {O_2}}$.
\label{fig2}}
\end{figure}

\begin{figure}

\plotone{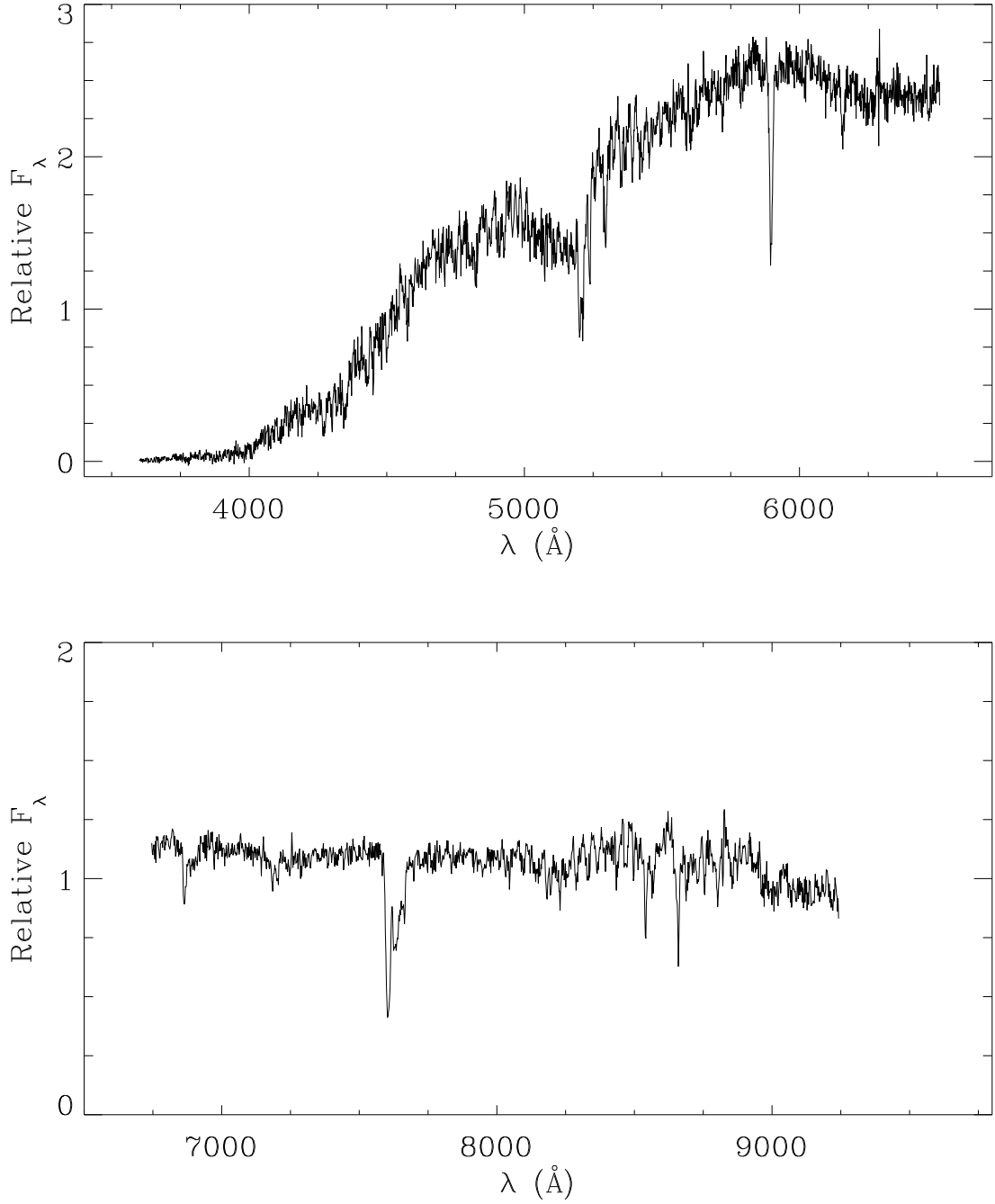}
\caption{Optical spectrum of the nearby background K star.  
The strength of the MgH feature centered around 5100 \AA{} may
indicate a metal poor atmosphere \citep{jac84,rei00}.
\label{fig3}}
\end{figure}

The lamp used to flat field the spectral images is itself spectrally
nonuniform and has a strong rise from $4500-7500$ \AA{}.\footnote
{http://www2.keck.hawaii.edu/realpublic/inst/lris/flats.html.}
Hence, the instrument and chip response were removed by flat fielding
but the remaining shape was a convolution of the stellar spectra and 
the ``flat'' lamp.  The background star appears to be a K dwarf whose 
exact spectral type cannnot be established from photometry because its
reddening is unknown and the low resolution spectrum precludes the
line measurements necessary for a determination.  Spectra extracted 
without flat fielding confirm that the flux of both GD392B and K star 
rises toward the dichroic cutoff near 6600 \AA{} (a fact corroborated 
by the photometry). Attempting to reconstruct the true shape of the 
continuua, the reduced spectra were multiplied by a blackbody with a 
temperature similar to that of the halogen flat lamp.  This
information was obtained from the manufacturer.\footnote
{http://www.oriel.com/tech/curves.htm.}  The resultant spectra
are shown in Figures \ref{fig2} and \ref{fig3}

\section{PROPER MOTION OF GD392A \& B}

A wide field, infrared proper motion survey for low mass
stellar and substellar companions to nearby white dwarfs is
being completed \citep{far03}.  Because the observations are
being conducted primarily at $J$ band ($1.25\mu$m), the survey
is particularly sensitive to cool objects such as late M, L and
early T type dwarfs.  A near infrared search is also sensitive
to cool white dwarfs.  This survey's sensitivity is slightly
less for cool degenerates which suffer collision induced
absorption (CIA) opacity at this wavelength relative to those
which do not.  In general, common proper motion companions
brighter than $J=19$ mag can be detected.

To measure proper motions and detect companions, GEOMAP is
used.  This is a standard program in the IRAF environment,
which generates a transformation between two sets of coordinates
corresponding to sources in the same field at two different epochs.
In this way proper motion stars can be identified and their
motions measured against the near zero motion of background
stars and galaxies, which provide a measure of the standard error.

\begin{figure}

\plotone{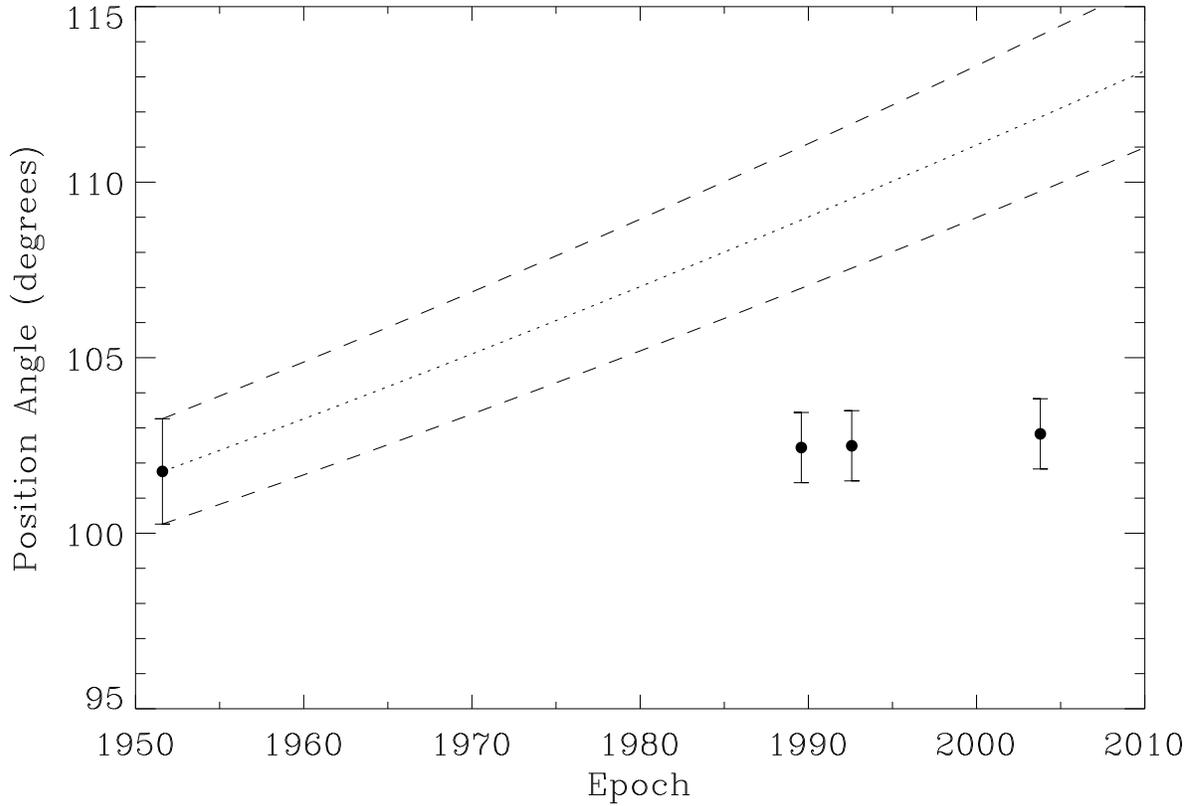}
\caption{Measured position angle of the GD392 system at
several epochs spanning 52 years.  Dashed lines bound a region
in which one would expect to find GD392B were it a stationary
background object.  The error in position angle includes
uncertainty in the difference of two centroids in a given
image plus uncertainty in the alignment accuracy of the image
itself.  The 2003 data point is from an infrared image and the
other three data points are from the digitized POSS I \& II
plates.
\label{fig4}}
\end{figure}

The common proper motion companion to GD392 (WD2058+342,
Greenstein 1984; Greenstein \& Liebert 1990, Wesemael et al.
1993) was detected in a routine examination of the digitized
POSS I \& II plates for confirming identity, proper motion and
coordinates of the primary for a finder chart prior to an
observing run.  A simple blinking of POSS I and POSS II frames
reveals the comoving companion at a separation of $46.2''$ and
a PA of $102.5^{\circ}$ (Figure \ref{fig1}).  Measurement of
proper motion between POSS epochs reveals that both GD392A
and B have the same proper motion over a 41 year baseline,
namely $\mu=0.17\pm0.01''$ ${\rm yr}^{-1}$ at $\theta=44\pm5^
{\circ}$. Furthermore, astrometric analysis of 2003 infrared
images confirms that the separation and PA between primary
and secondary have remained constant since 1951.  The USNO
B1.0 catalog \citep{mon03} gives a proper motion of $\mu=
0.168\pm0.002''$ ${\rm yr}^{-1}$ at $\theta=42.6\pm0.7^{\circ}$
for GD392 and no value is given for the companion.  \citet{gic65}
published a detected proper motion of $0.1''\leq\mu\leq0.2''$
${\rm yr}^{-1}$ at $45^{\circ}$ for the primary.  Both of
these measurements are consistent with the value presented
in this paper.  Figure \ref{fig4} shows the unchanging
position angle of the pair over several epochs.  

At a nominal distance of 50 pc (see \S4.2), the proper motion
of the pair gives $v_{\rm tan}=40$ km ${\rm s}^{-1}$ and thus there
is no reason to suspect this system is not a member of the local
disk population.

\section{RESULTS}
 
\subsection{\it Spectral Energy Distribution \& Temperature of GD392B}

Despite some relatively low SNR photometry on GD392B in
the near infrared, there are seven independent measurements
at these longer wavelengths, three of these at $2.2\mu$m
(Table \ref{tbl-1}).  In stark contrast with the similar optical
magnitudes and colors, GD392B appears roughly 1.0 magnitude
fainter than the nearby background K star at $1.2\mu$m, 2.0
magnitudes fainter at $1.6\mu$m, and 2.5 magnitudes fainter at
$2.2\mu$m in all the measurements.

The optical and infrared colors of GD392B point strongly
towards an ultracool white dwarf with strong CIA longward of
$1\mu$m.  No other stellar object could be so red in $V-I$ and
yet blue in $I-K$ \citep{ber95,han98a,har99,sau99,hod00,opp01}.
The optical spectrum confirms that GD392B is a featureless DC star
(Figure \ref{fig2}).  Hence GD392B must be a very cool white dwarf.

\begin{figure}

\plotone{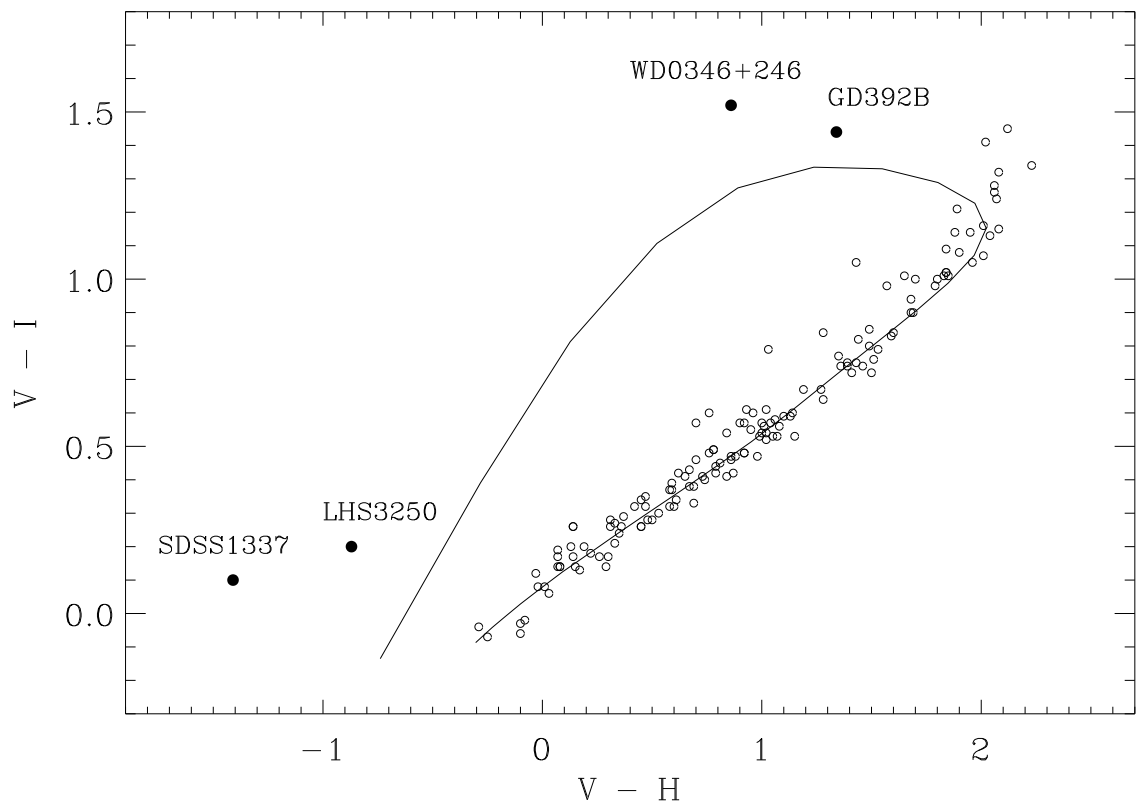}
\caption{$V-I$ vs. $V-H$ color-color diagram including all
known ultracool white dwarfs with published infrared photometry
(SDSS1337+00 does not have a $K$ band magnitude in the literature,
which is why $H$ is used here).  The solid line represents log
$g=8$ hydrogen atmosphere cooling tracks all the way down to 2000
K at lower left (P. Bergeron 2002, private communication).  The
turnoff corresponds to $T_{\rm eff}\sim4000$ K.  Open circles
represent cool white dwarfs from \citet{ber01b}.
\label{fig5}}
\end{figure}

Plotting GD392B on a $V-I$ vs. $V-H$ color-color diagram,
one sees that it is located beyond the turnoff for log $g=8.0$
pure hydrogen atmosphere white dwarfs, which corresponds to
$T_{\rm eff}\sim4000$ K (Figure \ref{fig5}).  There are only
three other ultracool white dwarfs in the literature for which
infrared photometry is available, WD0346+246, LHS3250 and
SDSS1337+00 \citep{har99,har01,hod00,ber02}.  GD392B's optical
colors are strikingly similar to those of WD0346+246 \citep{opp01}.  
Although they have almost identical optical colors, GD392B appears
to have more flux in the near infrared than does WD0346+246.  From
recent model atmosphere analyses of the three aforementioned stars, 
it is possible that GD392B is warmer, has different gravity, and/or 
contains a different ratio of hydrogen to helium in its atmosphere 
than does WD0346+246 \citep{ber01a,ber02}.

\begin{figure}

\plotone{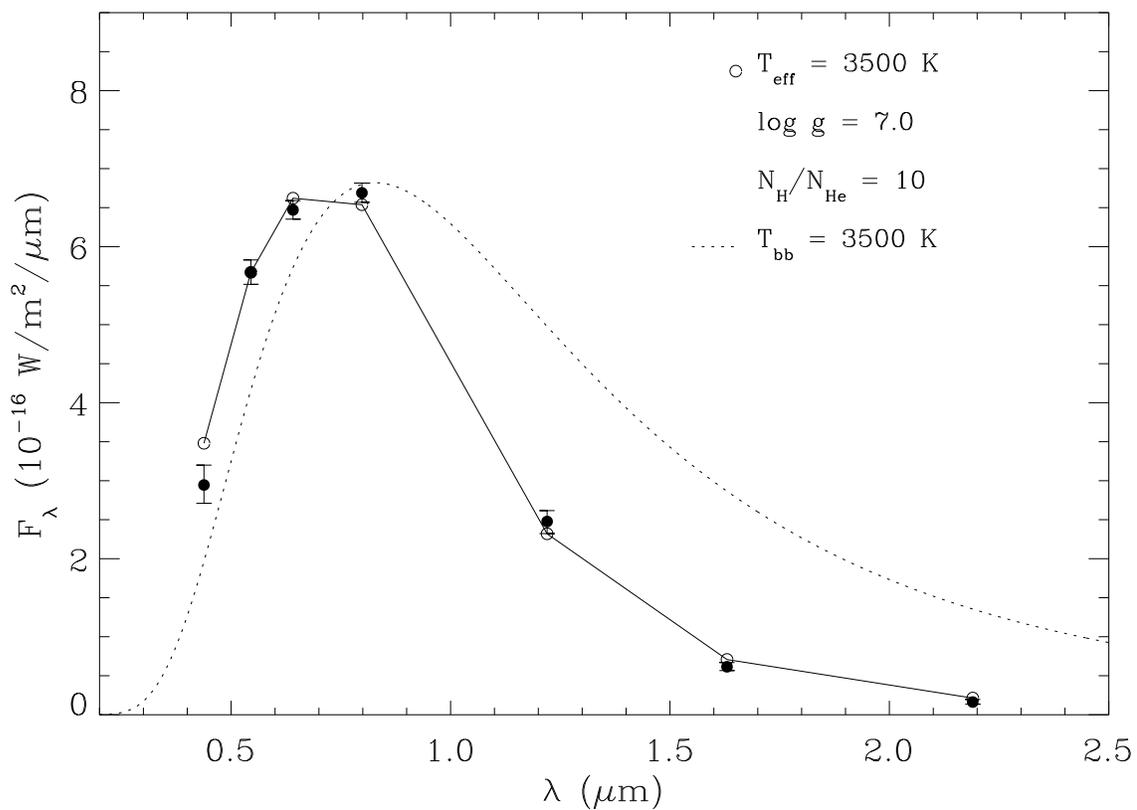}
\caption{Spectral energy distribution of GD392B as
determined from $BVRIJHK$ photometry.  Filled circles with
error bars represent the data.  Open circles and solid line
represent the mixed atmosphere model fit -- only the $B$ data
point is discrepant.  The dotted line is a blackbody of the
same temperature, demonstrating that the SED is reshaped by CIA.
\label{fig6}}
\end{figure}

The best way to estimate the temperature of GD392B is to fit
models to the $BVRIJHK$ photometry.  To evaluate its SED, the
magnitudes were converted into isophotal or average fluxes,
following the method of \citet{ber97}.  Different fits to the
data were tried using the pure hydrogen and mixed H/He atmosphere
model grids of P. Bergeron (2003, private communication).  As in
the case of WD0346+246 \citep{ber01a,opp01}, pure hydrogen and
mixed atmosphere models fail to reproduce the SED of GD392B in
detail.  And like \citet{opp01}, a low gravity solution appears
to provide a decent fit to the data.  In fact, every solution at
log $g>7.0$ seems to provide optical colors that are too blue
(too much CIA), or infrared colors that are too red (too little
CIA).  As shown in Figure \ref{fig6}, a decent fit is acheived at
log $g=7.0$, $T_{\rm eff}=3500$ K, and $N_{\rm H}/N_{\rm He}=10$.
Also shown in Figure \ref{fig6} is a blackbody of the same
temperature scaled to the peak flux of GD392B -- relative to
the blackbody, the SED of GD392B appears suppressed at
wavelengths longer than 8000 \AA{} and enhanced at shorter
wavelengths.

Because GD392B is a companion to a previously studied white
dwarf, it should be possible to deduce the distance to this
system and use it to constrain stellar radius $R$ and $M_V$.
One must keep in mind that the model predicted temperature of
GD392B will depend on its mass and atmospheric composition and
thus cannot be known exactly at present.  In \S4.2 a range of
possibilities is explored.

\subsection{\it Stellar Masses of the GD392 System} 

Using basic relations between luminosity and flux, one can
calculate a radius for GD392B if its distance and $T_{\rm eff}$
are known.  To obtain a distance, the $BVRIJHK$ photometry of the
primary (Table \ref{tbl-1}), GD392, was fitted with pure helium
model grids (P. Bergeron 2002, private communication) and its
measured SED was integrated directly.  The model fit provides
a $T_{\rm eff}$, and integrating the SED of GD392 yields a total 
flux.  A radius must be specified in order to calculate a distance 
to the primary.  The model fit to the SED of GD392 does not provide 
a radius determination because the colors of a DB star are largely 
unaffected by gravity.  Hence one is forced to make various 
assumptions for log $g$.  Table \ref{tbl-3} lists stellar parameters 
for GD392 with various gravities, and thus distances.  In the 
literature, GD392 has been noted as a DB5 \citep{gre84}, DBA?4 
\citep{gre90} and finally a DB5 again while explicitly stating 
``revised from \citet{gre90}'' \citep{wes93}.  The optical and 
infrared photometric data presented here are more consistent with 
a DB4 than a DB5.  Good agreement between model and measured colors 
is acheived at $T_{\rm eff}=11,625$ K (regardless of log $g$), which
actually corresponds to DB4.5.  This is the temperature used in all
calculations for the parameters of GD392.  For each distance estimate
to the GD392 system, there is an analytic constraint on
$R{T_{\rm eff}}^2$ for GD392B.  It is also critical that the
model predicted parameters, such as $M_V$, optical and infrared
colors corresponding to each combination of $R$ and $T_{\rm eff}$,
fit the photometric data on GD392B.

\begin{figure}

\plotone{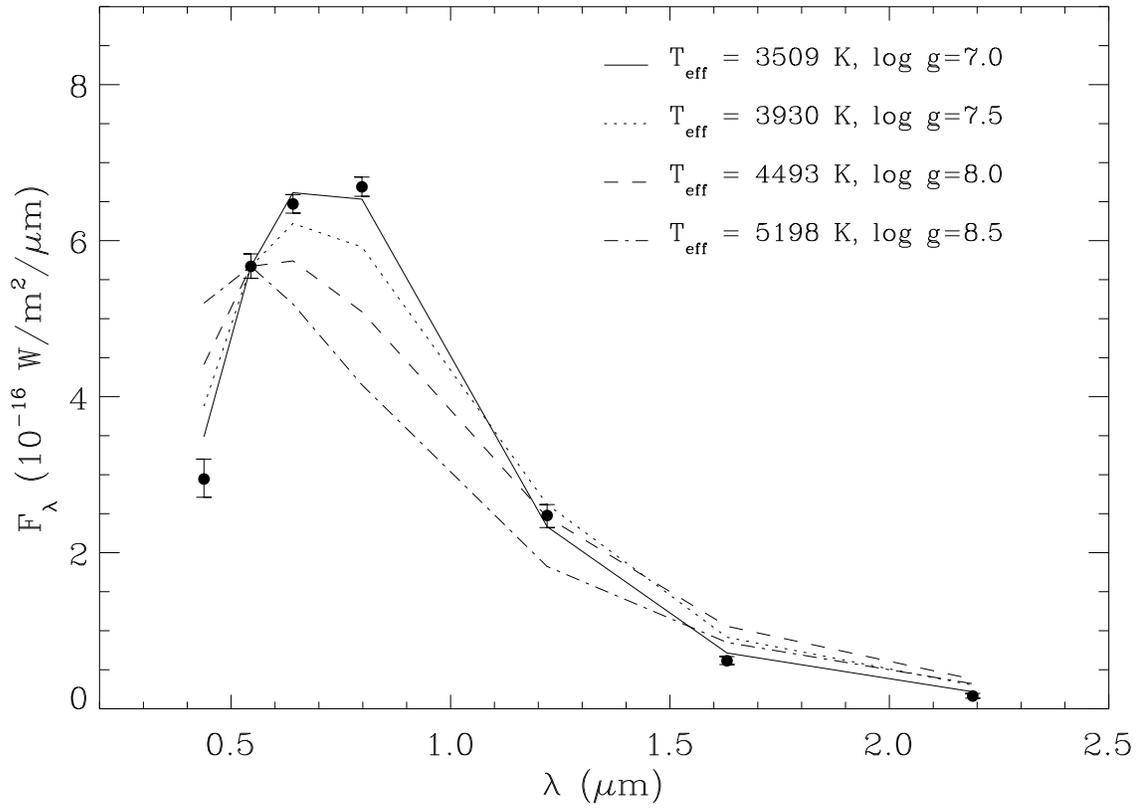}
\caption{Model fits to GD392B at 57.8 pc with
$N_{\rm H}/N_{\rm He}=10$.  All but the lowest gravity case
fail to reproduce the photometric measurements.
\label{fig7}}
\end{figure}

If GD392 is an average mass white dwarf (log $g=8.0$), a
solution is found almost identical to the preliminary fit in
Figure \ref{fig6}.  That is, at 57.8 pc GD392B must have a
temperature near 3500 K and a very large radius -- either an
unresolved binary or a low mass white dwarf.  This model fit
is consistent with the measured colors (excepting $B-V$, see
\S4.3) and matches the absolute magnitude of the secondary at
this distance; in fact, only a low gravity solution appears to 
provide such agreement between measured and predicted fluxes.
In Figure \ref{fig7}, four fits are shown satisfying the $M_V$
constraint at 57.8 pc for a range of gravities.  It is clear that
only the log $g=7.0$ fit comes near to matching the photometry.  
Although these fits have all been scaled to match the flux at
$0.55\mu$m, the obvious discrepancies of the higher gravity
solutions occur at any scaling; they predict too little flux in 
the optical or too much flux in the near infrared.  In addition, 
the log $g\geq7.5$ fits do not satisfy the $R{T_{\rm eff}}^2$ 
constraint at this distance.  This scenario is instructive for 
two reasons.  One is that there is no independent reason to 
suspect that GD392 is anything but an average mass white dwarf.  
In fact, the spectroscopic mass distribution of DB stars peaks 
at $0.59M_{\odot}$, with a very small dispersion of $0.06M_{\odot}$
\citep{bea96}.  The other reason is the failure of the higher 
gravity fits to match the data persists for any reasonable 
distance estimate.

\begin{figure}

\plotone{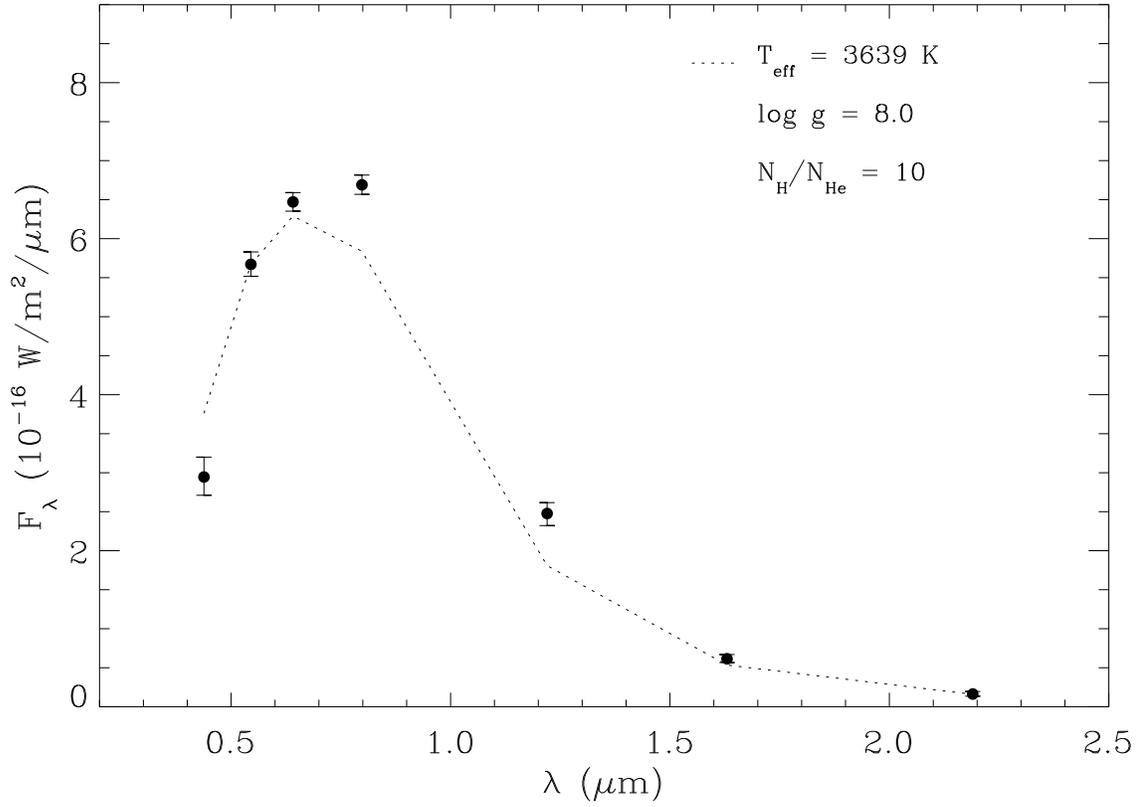}
\caption{Model fit to GD392B at 40.4 pc.  The failure of the
models to fit the data points persists for both higher gravities
and different amounts of photospheric helium.
\label{fig8}}
\end{figure}

Assuming the primary is a higher than average mass white
dwarf (log $g\geq8.5$) results in serious discrepancies between
model predicted and measured colors for GD392B.  This disagreement
appears to be insensitive to the type of atmosphere in the models.
Shown in Figure \ref{fig8} is a model fit for GD392B at 40.4 pc.  
Higher gravity solutions at this distance or closer are worse and 
are not shown for the sake of brevity.  Problems arise as well if 
one assumes the primary is a lower than average mass white dwarf 
(log $g\leq7.5$).  This leads to extreme low gravity solutions for
GD392B and the models used here do not include gravities below log
$g=7.0$.  Table \ref{tbl-4} lists all resulting fits for GD392B.  

While the possibility exists that GD392B is an unresolved
binary or helium core white dwarf, this is uncertain until the
distance to the system is known with sufficient precision and a
good model fit to its SED is achieved.  It seems very unlikely
that GD392B could have a such an unusually low mass as $0.15M_
{\odot}$, but it cannot be ruled out either.  It is conceivable
that a good model fit with a much higher $T_{\rm eff}$ (hence
predicting a smaller radius and a larger mass) could show good
agreement with the measured colors of GD392B.  All models used
here do not give such agreement.  A parallax measurement of the
primary is currently in progress (H. Harris 2003, private
communication).

\subsection{\it Atmospheric Composition of GD392B}

\begin{figure}

\plotone{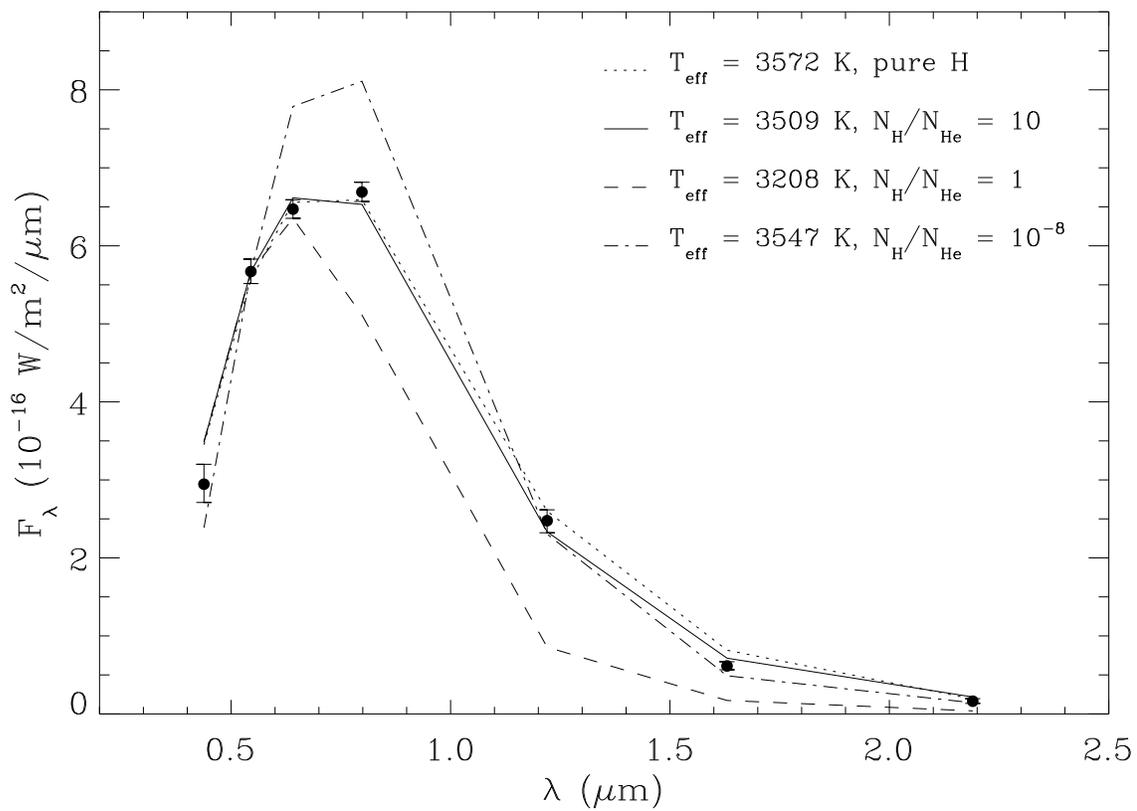}
\caption{Model fits to GD392B at 57.8 pc with log $g=7.0$.  
Only the $N_{\rm H}/N_{\rm He}=10$ and pure hydrogen models
have any success in reproducing the measurements.  Fits at
$N_{\rm H}/N_{\rm He}=10^{-1},10^{-2},10^{-3},10^{-5}$ have
been ommitted because they are more discrepant than those shown.
\label{fig9}}
\end{figure}

All attempts to fit the SED of GD392B with mixed H/He
atmosphere models having $N_{\rm H}/N_{\rm He}<10$ are
problematic.  In fact, for all temperatures and gravities,
the same problem mentioned previously occurs; colors too
blue in the optical and/or too red in the infrared to fit
the measured photometry, given the absolute magnitude
constraint.  The only possible exception being the pure
hydrogen models at log $g=7.0$.  Shown in Figure \ref{fig9}
are several model fits with varying ratios of H/He for the
lowest gravity case.  Only the pure hydrogen and the
$N_{\rm H}/N_{\rm He}=10$ models provide any kind of
decent fit to the data.  However, the pure hydrogen fit
has one minor fault when compared to the mixed atmosphere
fit -- it predicts an $M_V$ corresponding to 57.8 pc but does
not precisely fit the $R{T_{\rm eff}}^2$ constraint.  The
$N_{\rm H}/N_{\rm He}=10$, $T_{\rm eff}=3509$ K solution comes
very close to hitting all the data points except $B$ and
satisfies all constraints surprisingly well at 57.8 pc.  There
currently exists a discrepancy between model predicted and
measured $B$ band fluxes -- this wavelength is excluded from
fits to the energy distributions of very cool hydrogen
atmosphere white dwarfs \citep{ber97,ber01a,ber03}.  It is
noteworthy that GD392B appears to be the first ultracool white
dwarf for which a pure hydrogen atmosphere is not inconsistent
with the data.

At present there remains a general failure by model predictions
to fit the photometric observations of ultracool white dwarfs in
detail \citep{ber01a,ber02} -- a fact exemplified by GD392B.
Another case in point is the SED of WD0346+246, where both
\citet{opp01} and \citet{ber01a} found that a mixed atmosphere
model with an extremely low value of $N_{\rm H}/N_{\rm He}\sim10
^{-6}-10^{-9}$ provides a good fit to the data.  But because of
the unlikelihood and volatility of such a solution, \citet{ber01a}
introduced a pseudocontinuum opacity characterized by an ad hoc
damping function.  Although a good fit with $N_{\rm H}/N_{\rm He}
=0.77$ was achieved when the opacity was added to the models, it
is unclear whether this pseudocontinuum source plays a significant
physical role in the atmospheres of ultracool white dwarfs.  
Therefore, it is not yet possible to determine the atmospheric
parameters of the coolest degenerates with confidence.

It should be mentioned that with the possible exception
of the ultracool white dwarfs in Figure \ref{fig5}, all
suspected or confirmed low mass white dwarfs have hydrogen
rich atmospheres \citep{ber92,ber01b,ber02}.  Hence a model
fit to the photometry of GD392B with low mass and a mixed
H/He atmosphere is inconsistent with these findings.

\section{DISCUSSION}

GD392B is not the first or even the second ultracool white
dwarf for which an apparent large radius is possible or likely.  
LHS3250, WD0346+246 and SDSS1337+00 {\it all appear to be potentially
overluminous} \citep{har99,har01,opp01}.  In fact, the best fits
acheived by \citet{opp01} to WD0346+246 and F351-50 were with
log $g=6.5$, although for pure hydrogen atmospheres.  These
solutions are problematic because they predict temperatures
below 3000 K and such extreme low gravities.  Although unlikely, 
the present possibility that these ultracool degenerate stars
have anomalously low mass cannot be ruled out.  White dwarfs with
$M<0.45M_{\odot}$ have relatively large radii and are understood
to be the product of close binary evolution; their post main
sequence lifetimes cut short before helium burning begins
\citep{ber92,mar95,han98b}.  On the other hand, the potential
overluminosity of these stars can be explained equally well by
binary membership with a white dwarf companion of comparable
luminosity.  If this is correct, the GD392 system would then be
the second known triple degenerate.  In either case, binarity
is a strong possibility for most if not all four objects in Figure
\ref{fig5}, if the models are correct.

Is it possible that the discovery of these stars represents
a luminosity detection bias?  This would imply all known
ultracool degenerates were detected because they are relatively
bright compared to normal to high gravity single white dwarfs
with comparable temperatures.  In fact, the faintest white
dwarf known, ESO439-26 \citep{rui95}, has an extremely high
mass (log $g=9.0$), a temperature around 4500 K, and $V=20.5$
-- just above the detection limit of very large sky surveys like
the SERC/ESO survey in the southern hemisphere and the Palomar
survey (POSS II) in the north \citep{rei91}.  What if ESO439-26
had $T_{\rm eff}=3500$ K instead of $\sim4500$ K?  Would it have
still been detected?  Almost certainly not, since models predict
it would have been too faint at $V=22.5$ and $B-V=1.0$.  A similar
calculation was performed for GD392B, assuming it is a 3500 K mixed
H/He atmosphere degenerate with log $g=7.0$ at 57.8 pc.  Would it
still have been detected were it a higher mass white dwarf with
log $g=8.5$?  Likely not at $V=21.0$ and $B-V\sim1.2$ which is
just beyond the POSS II red plate limit and right at the blue
plate limit.  It probably would not have been detected in the
infrared at $J=19.2$.  Similar conclusions can be drawn for
WD0346+246 at $V=19.1$, but possibly not LHS 3250 at $V=18.0$.  
If we similarly push back the magnitude of SDSS1337+00 to
$V=20.8$ it would still be detectable in the Sloan survey but
not on the POSS II plates.

The claim that all known ultracool white dwarfs are
overluminous is not being made here.  It is simply being
stated that {\it if} the majority of them are indeed brighter
than expected due to low gravity or binarity, then all surveys
with the exception of Sloan have been biased against detecting
average to higher mass single white dwarfs at the very bottom of
the disk cooling sequence.  In fact, one should expect that the
oldest white dwarfs -- those which spent very little time on the
main sequence -- would have come from fairly massive progentiors.  
Furthermore, the initial mass to final mass relation for white
dwarfs indicates that higher mass main sequence stars tend to form 
higher mass degenerates \citep{wei87,wei90,wei00,bra95}.  This
implies that the oldest and least luminous white dwarfs should 
have high masses, small radii, and be up to 2.4 magnitudes fainter
than a low mass white dwarf with the same temperature.  Hence there
is a possibility that the lowest luminosity disk white dwarfs
have yet to be detected.

Of course the other possibility (and, according to Occam's
razor, the most likely) is simply a failure of the current
models to explain the photometric observations.  Also, in this
case, the data do not constrain the mass of GD392 and therefore
we do not know distance to the system.  Despite the fact that
the model predictions fail to reproduce the colors of GD392B in
all but the log $g=7.0$ case, it is certainly possible that it
has a higher mass.  This cannot be ruled out and future
improvements in models may result in the disappearance of
the overluminosity issue. 

\section{CONCLUSION}

Spectroscopy and photometry confirm a cool DC white dwarf
common proper motion companion to GD392.  It is the first
ultracool white dwarf ($T_{\rm eff}<4000$ K) to be discovered
as a companion to another star.  Although the primary is a well
studied DB white dwarf, GD392 does not have a mass estimate, nor
any kind of reliable parallax.  In fact, its helium lines are
very weak \citep{wes93} and therefore it is not possible to
spectroscopically determine its mass.  Neither is it possible
to photometrically determine its mass because the optical and
infrared colors of relatively hot helium atmosphere white dwarfs
are insensitive to gravity.  Therefore the distance to the GD392
system is not well constrained and the parameters we derive
for GD392B must come from a range of assumed values for the
mass of the primary.

Despite this difficulty, it appears that all model fits
fail to reproduce the SED of GD392B with the exception of a
solution with log $g=7.0$, $T_{\rm eff}=3509$ K, and $N_{\rm H}
/N_{\rm He}=10$ (or similar values with pure hydrogen).  It is
the only set of parameters for which the model predicted fluxes
fall within $1-1.5\sigma$ of all but one photometric data point
and which satisfies all constraints.  However, the model fit alone
is not sufficient to conclude that GD392B is overluminous.  It is
possible that current models simply cannot reproduce the observed
spectral energy distribution of GD392B with a normal to higher
surface gravity, but perhaps ongoing and future improvements will.
Based on the data and the models used here, along with the
published analyses of similar objects, GD392B may be an unresolved
binary with a companion of comparable luminosity or is itself a
low mass white dwarf.  Albeit unlikely, this could also be the
case for other known ultracool degenerates.  If correct, this may
hold interesting consequences  for the population of stars at the
faint end of the local disk white dwarf luminosity function.

\acknowledgments

Some of the data presented herein were obtained at the Keck
Observatory, which is operated as a scientific partnership
among the California Institute of Technology, the University of
California and the National Aeronautics and Space Administration
(NASA).  The author would like to extend his sincere gratitude to
C. Steidel, D. Erb and N. Reddy for obtaining the optical spectrum
of GD392B and to E. E. Becklin \& I. Song for making the NIRC
observations.  Acknowledgement goes to E. E. Becklin \& B. Zuckerman
for careful readings of the manuscript and to B. Hansen for many
helpful conversations.  P. Bergeron kindly provided his models for
use in this work.  Thanks to the referee N. Hambly for very
helpful suggestions and comments.  A special thanks goes to C. 
McCarthy for donating his time and computing skills much to the
author's benefit.  The digitized version of the POSS I \& II plates
were provided as a service by the Space Telescope Science Insititute.
This research has been supported in part by grants from NASA to
UCLA.

\clearpage

\clearpage

\clearpage

\clearpage 

\clearpage

\clearpage

\clearpage

\clearpage

\clearpage

\clearpage

\begin{deluxetable}{lcccc}
\tablecaption{Photometric Data for the GD392 System. \label{tbl-1}}
\tablewidth{0pt}
\tablehead{
\colhead{Band} 				&
\colhead{$\lambda_{0}$ ($\mu$m)}	&
\colhead{GD392A (mag)}			&
\colhead{GD392B (mag)}			&
\colhead{K star (mag)}}

\startdata

$B$	&0.44	&15.71	&$20.82\pm0.09$	&$20.52\pm0.07$ \\
	&	&15.79	&$20.74\pm0.06$	&$20.51\pm0.06$	\\

$V$	&0.55	&15.67	&$19.50\pm0.03$	&$19.31\pm0.03$ \\
	&	&15.69	&$19.59\pm0.04$	&$19.44\pm0.04$	\\

$R$ 	&0.64	&15.61  &$18.80\pm0.02$	&$18.55\pm0.02$ \\
	&	&15.63	&$18.83\pm0.04$	&$18.54\pm0.04$	\\

$I$ 	&0.80	&15.62	&$18.06\pm0.02$	&$17.73\pm0.02$	\\
	&	&15.69	&$18.18\pm0.04$	&$17.81\pm0.04$ \\

$J$	&1.25	&15.73	&$17.73\pm0.07$	&$16.79\pm0.06$ \\
	&	&15.77	&$17.64\pm0.06$	&$16.75\pm0.05$ \\

$H$ 	&1.65	&15.79 	&$18.16\pm0.09$	&$16.16\pm0.05$ \\
	&	&15.81	&$17.93\pm0.10$	&$16.17\pm0.05$ \\

$K^{'}$	&2.12	&15.86	&$18.49\pm0.25$	&$16.09\pm0.07$ \\
	&	&15.88	&$18.64\pm0.29$	&$16.10\pm0.06$ \\

$K$	&2.21	&\nodata&$18.53\pm0.20$	&$15.99\pm0.05$	\\

\enddata

\tablecomments{Photometric uncertainties for GD392A are 3\%.
The first values listed for each filter are the primary data
set used in the analysis for GD392B.  All other data are
secondary (\S2.1).}

\end{deluxetable}

\clearpage

\begin{deluxetable}{lccc}
\tablecaption{Adopted Magnitudes and Fluxes for GD392B. \label{tbl-2}}
\tablewidth{0pt}
\tablehead{
\colhead{Band} 				&
\colhead{$\lambda_{0}$ ($\mu$m)}	&
\colhead{Magnitude}			&
\colhead{$F_{\lambda}$ ($10^{-16}$ W/${\rm m}^2$/$\mu$m)}}

\startdata

$B$		&0.44		&$20.82\pm0.09$	&$2.95\pm0.24$	\\
$V$		&0.55		&$19.50\pm0.03$	&$5.67\pm0.16$	\\
$R$ 		&0.64		&$18.80\pm0.02$	&$6.47\pm0.12$	\\
$I$ 		&0.80		&$18.06\pm0.02$	&$6.69\pm0.12$	\\
$J$		&1.22		&$17.73\pm0.07$	&$2.47\pm0.15$	\\
$H$ 		&1.63		&$18.16\pm0.09$	&$0.62\pm0.05$	\\
$K$		&2.19		&$18.51\pm0.23$	&$0.16\pm0.03$	\\

\enddata

\tablecomments{$BVRI$ is on the Johnson-Cousins system and
$JHK$ is on the Johnson-Glass system, collectively known as the
Johnson-Cousins-Glass system \citep{bes88,bes90}.  Conversions
between near infrared filter sets were ignored (\S2.1).}

\end{deluxetable}

\clearpage

\begin{deluxetable}{lrrr}
\tablecaption{Possible Stellar Parameters for GD392A. \label{tbl-3}}
\tablewidth{0pt}
\tablehead{
\colhead{$T_{\rm eff}=11,625$ K} 	&
\colhead{log $g$ = 8.0}			&
\colhead{log $g$ = 8.5}			&
\colhead{log $g$ = 9.0}}

\startdata
$d$ (pc)		&57.8		&40.4		&25.9		\\
$M/M_{\odot}$		&0.59		&0.90		&1.19		\\
$R/R_{\odot}$		&0.0127		&0.0089		&0.0057		\\
$M_V$ 			&11.87		&12.65		&13.61		\\
$M_{\rm bol}$ 		&11.20		&11.98		&12.93		\\
log $(L/L_{\odot})$ 	&-2.58		&-2.89		&-3.27		\\
Cooling Age (Gyr)	&0.45	 	&0.91		&1.78		\\
 
\enddata

\tablecomments{Assuming log $g\leq7.5$ for the primary
leads to unphysical solutions for the secondary.}

\end{deluxetable}

\clearpage

\begin{deluxetable}{lrrr}
\tablecaption{Possible Stellar Parameters for GD392B. \label{tbl-4}}
\tablewidth{0pt}
\tablehead{
\colhead{} 				&
\colhead{$d=57.8$ pc} 			&
\colhead{$d=40.4$ pc\tablenotemark{\dag}} 	&
\colhead{$d=25.9$ pc\tablenotemark{\dag}}}

\startdata

$T_{\rm eff}$ (K) 	&3509		&3639		&3875		\\
log $g$			&7.00		&7.95		&8.72		\\
$M/M_{\odot}$		&0.153		&0.576		&1.066		\\
$R/R_{\odot}$		&0.0205		&0.0133		&0.0075		\\
$M_V$			&15.69 		&16.47		&17.43		\\
$M_{\rm bol}$ 		&15.35		&16.14		&17.27		\\
log $(L/L_{\odot})$ 	&-4.24		&-4.56		&-4.94		\\
Cooling Age (Gyr)	&2.92	 	&9.92		&8.62		\\
 
\enddata

\tablecomments{No satisfactory model fit is found for anything
but a low surface gravity white dwarf at approximately 58 pc
(see \S4.2).}

\tablenotetext{\dag}{These solutions show large
discrepancies with the measured colors of GD392B but have
been included for the sake of completeness.}

\end{deluxetable}

\end{document}